\documentclass[aps,prl,twocolumn,nofootinbib]{revtex4-1}

\usepackage{amsmath}
\usepackage{multirow}

\newcommand{\beq}{\begin{equation}}                                             
\newcommand{\eeq}{\end{equation}}     
\newcommand{\bea}{\begin{eqnarray}}                                             
\newcommand{\eea}{\end{eqnarray}}     

\usepackage{graphicx}

\begin{document} 

\preprint{MS-TP-20-28}

\title{Absolute neutrino mass as the missing link to the dark sector}

\author{Thede de Boer$^a$}
\author{Michael Klasen$^a$}
\author{Caroline Rodenbeck$^b$}
\author{Sybrand Zeinstra$^a$}

\affiliation{$^a$ Institut f\"ur Theoretische Physik, Westf\"alische Wilhelms-Universit\"at M\"unster, Wilhelm-Klemm-Stra\ss{}e 9, 48149 M\"unster, Germany}
\affiliation{$^b$ Institut f\"ur Kernphysik, Westf\"alische Wilhelms-Universit\"at M\"unster, Wilhelm-Klemm-Stra\ss{}e 9, 48149 M\"unster, Germany}

\begin{abstract}
  With the KATRIN experiment, the determination of the absolute neutrino mass scale
  down to cosmologically favored values has come into reach. We show that this
  measurement provides the missing link between the Standard Model and the dark
  sector in scotogenic models, where the suppression of the neutrino masses is
  economically explained by their only indirect coupling to the Higgs field. We
  determine the linear relation between the electron neutrino mass and the scalar
  coupling $\lambda_5$ associated with the dark neutral scalar mass splitting to be
  $\lambda_5=3.1\times10^{-9}\ m_{\nu_e}/$eV. This relation then induces correlations
  among the dark matter (DM) and new scalar masses and their Yukawa couplings.
  Together, KATRIN and
  future lepton flavor violation experiments can then probe the fermion DM
  parameter space, irrespective of the neutrino mass hierarchy and CP phase.
\end{abstract}

\maketitle

\section{Introduction}
\label{sec:1}

\vspace*{-90mm}
MS-TP-20-28
\vspace*{85mm}

The identification of cold Dark Matter (DM) -- a mysterious particle that according
to most cosmological models is five times more abundant in the Universe than
ordinary matter -- is one of the most urgent challenges in modern physics. Neutrinos
as only weakly interacting massive particles in the Standard Model (SM)
have the right characteristics of a DM candidate, but are neither cold nor
can they, due to their tiny masses, contribute more than a small fraction (between
0.5 and 1.6\%) to the measured total DM relic density \cite{Tanabashi:2018oca,%
  Klasen:2015uma}. Nevertheless the idea that neutrinos and DM might be related is
intriguing and has led to an enormous theoretical activity on so-called radiative
seesaw models, where the suppression of the SM neutrino masses is due to their only
indirect interaction (via DM) with the SM Higgs field \cite{Ma:2006km,Restrepo:2013aga,%
Cai:2017jrq}.

While the fact that at least two of the three neutrino flavors are massive has been
deduced about 20 years ago from atmospheric \cite{Fukuda:1998mi} and solar
\cite{Ahmad:2001an,Ahmad:2002jz} neutrino oscillations, their absolute masses are
still unknown. The KATRIN experiment has recently improved the upper limit on the
electron (anti)neutrino mass to 1.1 eV \cite{Aker:2019uuj} and ultimately aims at
a sensitivity of 0.2 eV \cite{Drexlin:2013lha}. This value would rival the cosmological
constraint on the sum of the SM neutrino masses of $\sum_i m_{\nu_i} < 0.12$ eV,
assuming the $\Lambda$CDM model and normal hierarchy (NH), the minimal value
allowed by oscillation experiments being 0.06 eV \cite{Vagnozzi:2017ovm,Aghanim:2018eyx}. An inverted
hierarchy (IH) is still a possiblity, although the long-baseline experiments T2K and
NO$\nu$A and further evidence from reactor and atmospheric neutrinos favor NH.
For the CP phase, T2K and to a lesser degree also NO$\nu$A data seem to favor
$3\pi/2$ ($\pi/2$) for NH (IH) \cite{Esteban:2018azc}.

\begin{figure}[t]
\begin{center}
\includegraphics[width=0.3\textwidth]{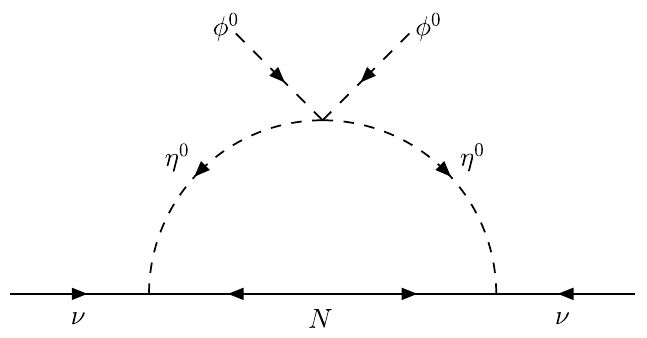}
\caption{Neutrino mass generation in scotogenic models.}
\label{fig:01}
\end{center}
\end{figure}

In radiative seesaw models, the SM neutrinos $\nu$, even under a discrete $Z_2$
symmetry, interact with the (also $Z_2$-even) SM Higgs field $\phi$
and obtain their masses via a dark ($Z_2$-odd) sector, which contains only a small number of new
multiplets (typically up to four new scalar/fermion singlets, doublets or triplets
under SU(2)$_L$) \cite{Restrepo:2013aga}. In Ma's famous scotogenic model (see Fig.\
\ref{fig:01}),
only one additional scalar doublet $\eta$ and (for three massive SM neutrinos)
three generations of fermion singlets $N_i$ (sterile neutrinos with $i=1,2,3$) are
required \cite{Ma:2006km}. The parameter space is therefore much smaller than, e.g.,
the one of supersymmetry and can be better constrained with neutrino oscillation
data via the Casas-Ibarra method \cite{Casas:2001sr}, limits on lepton flavor
violation (LFV) \cite{Toma:2013zsa}, and measurements of the DM relic density
\cite{Vicente:2014wga}. Nevertheless, these previous analyses found that the dark
scalar/fermion masses as well as their scalar and Yukawa couplings could still
vary over several orders of magnitude.

In this Letter, we demonstrate that a determination of the absolute electron
neutrino mass, which has now come into reach, will provide additional stringent
constraints on the dark sector of the scotogenic model in a way that is almost
independent of the neutrino hierarchy and CP phase. In particular, we determine
the linear relation between the absolute electron neutrino mass and the scalar
coupling associated with the mass splitting of the dark neutral scalars. This
linear dependence induces correlations among the other parameters of the model,
i.e.\ the DM and scalar masses and their Yukawa couplings, which we can also
quantify. Together, current neutrino mass and future LFV experiments can then
probe almost the entire fermion DM parameter space.

\section{The scotogenic model}
\label{sec:2}

In the original scotogenic model, the SM is enlarged with a dark sector containing
only two new types of fields, a complex Higgs doublet $\eta$ and three generations
of fermion singlets $N_i$ \cite{Ma:2006km}. In the Lagrangian of this model
\beq
 \mathcal{L}_N=-\frac{m_{N_i}}{2}N_iN_i+y_{i\alpha}(\eta^\dagger L_\alpha) N_i
+\mathrm{h.c.}-V,
\eeq
we define fermions in terms of Weyl spinors and denote the three generations of
left-handed SM lepton doublets with $L_\alpha$ ($\alpha=1,2,3$). The fermion singlet
with the smallest mass $m_{N_i}$ is assumed to be the DM candidate. SM neutrinos
$\nu$ couple to the SM Higgs field $\phi$ and obtain their mass only at one loop
(see Fig.\ \ref{fig:01}) via the $3\times3$ matrices of Yukawa couplings
$y_{i\alpha}$ and the scalar potential
\bea
V&=&
m_{\phi}^2\phi^\dag\phi+m_\eta^2\eta^\dag\eta
+\!\frac{\lambda_1}{2}\left(\phi^\dag\phi\right)^2
\!+\!\frac{\lambda_2}{2}\left(\eta^\dag\eta\right)^2
\!+\lambda_3\left(\phi^\dag\phi\right)\nonumber\\
&&\left(\eta^\dag\eta\right)
+\lambda_4\left(\phi^\dag\eta\right)\left(\eta^\dag\phi\right)
+\frac{\lambda_5}{2}\left[\left(\phi^\dag\eta\right)^2
+\left(\eta^\dag\phi\right)^2\right] .
\eea

The parameters $m_\phi$ and $\lambda_1$ are fixed by the known SM Higgs vacuum
expectation value (VEV) $\langle\phi^0\rangle=246$ GeV$/\sqrt{2}$ \cite{Tanabashi:2018oca}
and the LHC measurement of the (squared) SM Higgs boson mass $m_h^2 = 2 \lambda_1
\langle\phi^0 \rangle^2 = - 2 m_\phi^2 = (125$ GeV$)^2$ \cite{Aad:2015zhl}. To ensure
that the scalar potential is bounded from below and the vacuum is stable, we must have
\begin{equation}
\begin{gathered}
 \lambda_1 > 0,\
 \lambda_2 > 0,\
 \lambda_3 > - \sqrt{\lambda_1 \lambda_2},\\
 \lambda_3 + \lambda_4 - |\lambda_5|>- \sqrt{\lambda_1 \lambda_2},
\end{gathered}
\end{equation}
while perturbativity imposes $|\lambda_{2,3,4,5}|<4\pi$. The inert doublet $\eta$ does
not acquire a VEV, so that $\lambda_2$ induces only self-interactions and decouples
from the phenomenology. We set $\lambda_2=0.5$ without loss of generality.
The masses of the charged scalar component $\eta^+$ and the real and imaginary parts of
the neutral component $\eta^0=(\eta_R+i\eta_I)/\sqrt{2}$ are then
\bea
m_{\eta^+}^2&=&m_\eta^2+\lambda_3\langle\phi^0\rangle^2,\nonumber\\
m_R^2&=&m_{\eta}^2+\left(\lambda_3+\lambda_4+\lambda_5\right)\langle\phi^0\rangle^2,\\
m_I^2&=&m_{\eta}^2+\left(\lambda_3+\lambda_4-\lambda_5\right)\langle\phi^0\rangle^2.\nonumber
\eea
Note that it is natural for $\lambda_5$ and the mass difference $m_R^2-m_I^2=2\lambda_5
\langle\phi^0\rangle^2$ to be small, since if $\lambda_5$ was exactly zero, it would
induce a conserved lepton number and massless neutrinos \cite{Kubo:2006yx}. Following
previous work \cite{Vicente:2014wga}, we scan over the range $10^{-12}< |\lambda_5|<10^{-8}$.
For vanishing $\lambda_3$, the LEP limit on charged particles \cite{Abbiendi:2003yd}
implies a lower limit on the scalar mass range $m_\eta\in[0.1;10]$ TeV, which we
also employ for the sterile neutrino masses $m_{N_i}$. As $m_\eta^2$ dominates over
$\langle\phi_0\rangle^2$ in much of the parameter space, the scalar couplings
$\lambda_3$ and $\lambda_4$ will play a subdominant role, and $\eta^+$ will be close
in mass to both $\eta_R$ and $\eta_I$.

\section{Experimental constraints}
\label{sec:3}

The SM neutrino mass matrix $(m_{\nu})_{\alpha\beta}=(y^{T}\Lambda y)_{\alpha\beta}$
can be written in a compact form using the Yukawa matrix $y$ and the diagonal mass
matrix $\Lambda$ with eigenvalues
\beq
\Lambda_i=\frac{m_{N_i}}{32\pi^2}
\left[\frac{m_R^2}{m_R^2-m_{N_i}^2}\log\left(\frac{m_R^2}{m_{N_i}^2}\right)-
  (R\rightarrow I)
  \right] .
\eeq
It is diagonalized by the PMNS matrix $U$,
\beq
 U^{T} \, m_{\nu} \, U=\hat{m}_{\nu}\equiv \text{diag}(m_1,m_2,m_3).
\eeq
This implies that, for a given set of masses in $\Lambda_i$, the Yukawa couplings
\beq
 y=\sqrt{\Lambda}^{-1}R\sqrt{\hat{m}_\nu}U^{\dag}
\eeq
can be constrained, up to an orthogonal matrix $R$ depending on three arbitrary rotation
angles $\theta_i\in[0;2\pi]$ \cite{Casas:2001sr}, on top of the perturbativity bound
$|y_{i\alpha}|^2<4\pi$ and vacuum stability requirements \cite{Lindner:2016kqk}, by
the measured neutrino mass
differences and mixing angles, which we apply at $3\sigma$  \cite{Esteban:2018azc}.
When $\lambda_5\ll1$ and $m_R^2 \approx m_I^2$, the SM neutrino
mass matrix simplifies to
\bea
(m_\nu)_{\alpha\beta}&\approx&2\lambda_5 \langle\phi^0\rangle^2
\sum_{i=1}^3\frac{y_{i\alpha} y_{i\beta}m_{N_i}}{32\pi^2(m_{R,I}^2-m_{N_i}^2)}\nonumber\\
 &\times& \left[1+\frac{m_{N_i}^2}{m_{R,I}^2-m_{N_i}^2}
\log\left(\frac{m_{N_i}^2}{m_{R,I}^2}\right)\right],
\label{eq:8}
\eea
i.e.\ it is not only bilinear in $y$, but also linear in $\lambda_5$.

We then impose current (and study the constraining power of future) upper bounds
on the most important LFV processes
\bea
{\rm BR}(\mu \to e \gamma)&<&4.2\cdot10^{-13}\ \mbox{\cite{TheMEG:2016wtm}}\ (2\cdot10^{-15}\ \mbox{\cite{Renga:2018fpd}}),\nonumber\\
{\rm BR}(\mu \to 3e )&<&1.0\cdot10^{-12}\ \mbox{\cite{Bellgardt:1987du}}\ (10^{-16} \mbox{\cite{Blondel:2013ia}}), \\
{\rm CR}(\mu - e,{\rm Ti})&<&4.3\cdot10^{-12}\ \mbox{\cite{Dohmen:1993mp}}\ (10^{-18}\ \mbox{\cite{PRIME}}).\nonumber 
\eea
The branching ratios (BRs) and conversion rates (CRs) depend on the charged scalar
and sterile neutrino masses and their Yukawa couplings through dipole/non-dipole form
factors and box diagrams \cite{Toma:2013zsa} and are calculated with SPheno 4.0.3
\cite{Porod:2011nf}.

We also restrict the DM relic density with micrOMEGAs 5.0.8
\cite{Belanger:2018ccd} to the central value $\Omega h^2=0.12$ measured by Planck
\cite{Aghanim:2018eyx}, allowing for a theoretical uncertainty of $0.02$
\cite{Harz:2016dql,Branahl:2019yot}. In the standard freeze-out scenario, the
relic density results from DM annihilation processes in the early Universe, here
of the lightest sterile neutrinos into leptonic final states via charged and
neutral scalars in the $t$-channel. Coannihilation processes, which may occur in
fine-tuned scenarios with nearly mass-degenerate scalars and fermions
\cite{Klasen:2013jpa}, are required to contribute less than 1\%. Direct DM
detection is theoretically possible at one loop, but is currently beyond the
experimental reach \cite{Ibarra:2016dlb}.

\section{Numerical results}
\label{sec:4}

\begin{figure}[t]
\begin{center}
\includegraphics[width=0.49\textwidth]{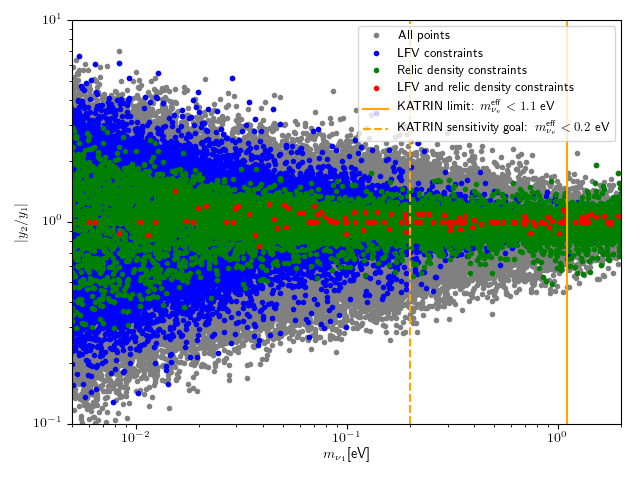}
\caption{Ratio of Yukawa couplings as a function of the lightest neutrino mass $m_{\nu_1}$
  with mass
  difference/mixing (grey), LFV (blue), relic density (green) and all constraints (red
  points).}
\label{fig:02}
\end{center}
\end{figure}

For electron neutrino masses of 1.1 eV to 0.2 eV as currently explored by KATRIN
\cite{Aker:2019uuj,Drexlin:2013lha}, i.e.\ larger than the minimal $\sum_i m_{\nu_i} >
0.06$ eV, but approaching the cosmological upper limit of 0.12 eV \cite{Vagnozzi:2017ovm,Aghanim:2018eyx},
the mass differences, PMNS matrix $U$ and rotation angles $\theta_i$ play a subdominant
role, and the eigenvalues of the Yukawas matrices $y_\alpha$ take similar values. This is
demonstrated in Fig.\ \ref{fig:02} (grey points), where the ratio $|y_2/y_1|$ varies
over its full range at low $m_{\nu_1}$, but only by about a factor of two at large
$m_{\nu_1}$. In addition, the LFV processes $l_\alpha\to l_\beta\gamma$ and $l_\alpha\to
3l_\beta$ impose upper limits on both $y_\alpha$ and $y_\beta$, limiting their ratio
further (blue). Conversely, to obtain the correct DM relic density, the Yukawas must
not be too small (green), so that the combination of all constraints leads indeed
to $|y_2/y_1|\sim1$ (red points). We have verified that this result is independent of the
neutrino mass hierarchy and holds also for the ratios $|y_3/y_1|$ and $|y_3/y_2|$.

\begin{figure}[t]
\begin{center}
\includegraphics[width=0.49\textwidth]{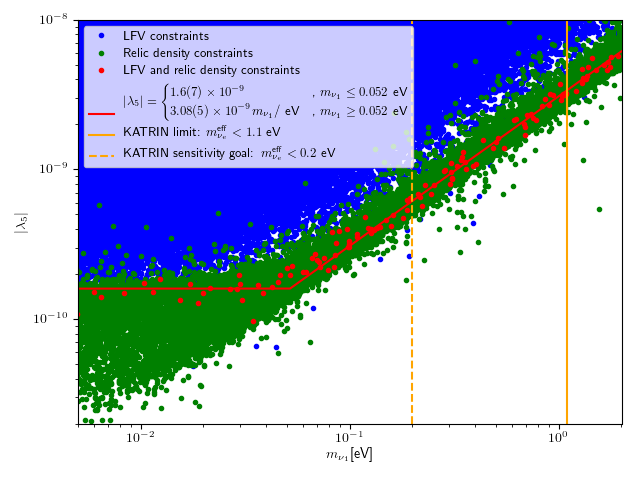}
\caption{The dark sector-Higgs boson coupling $|\lambda_5|$ as a function of lightest
  neutrino mass $m_{\nu_1}$
  with LFV (blue), relic density (green) and all constraints (red points).}
\label{fig:03}
\end{center}
\end{figure}

The linear dependence of the neutrino mass matrix $(m_\nu)_{\alpha\beta}$ on the
dark sector-Higgs boson coupling $|\lambda_5|$ in Eq.\ (\ref{eq:8}) can then be made
explicit by studying the dependence of $|\lambda_5|$ on the lightest eigenvalue
$m_{\nu_1}$. It emerges in Fig.\ \ref{fig:03} after imposing LFV
(blue), relic density (green) and all (red points) constraints and can be fitted at 90\% C.L.\ to
\beq
|\lambda_5|=\left\{\begin{array}{ll}
(3.08\pm0.05)\times10^{-9} \ m_{\nu_1}/{\rm eV} & ({\rm NH})\\
(3.11\pm0.06)\times10^{-9} \ m_{\nu_1}/{\rm eV} & ({\rm IH})\end{array}\right.,
\eeq
the sign being arbitrary, while below $m_{\nu_1}=0.052$ eV the heaviest neutrino
mass dominates and
\beq
|\lambda_5|=\left\{\begin{array}{ll}
(1.6\pm0.7)\times10^{-10} & ({\rm NH})\\
(1.7\pm1.5)\times10^{-10} & ({\rm IH})\end{array}\right.\hspace*{13.5mm}
\eeq
becomes independent of $m_{\nu_1}$. The dark sector-Higgs boson coupling $\lambda_5$
can therefore be predicted, once the absolute neutrino mass scale is known.

\begin{figure}[b]
\begin{center}
\includegraphics[width=0.49\textwidth]{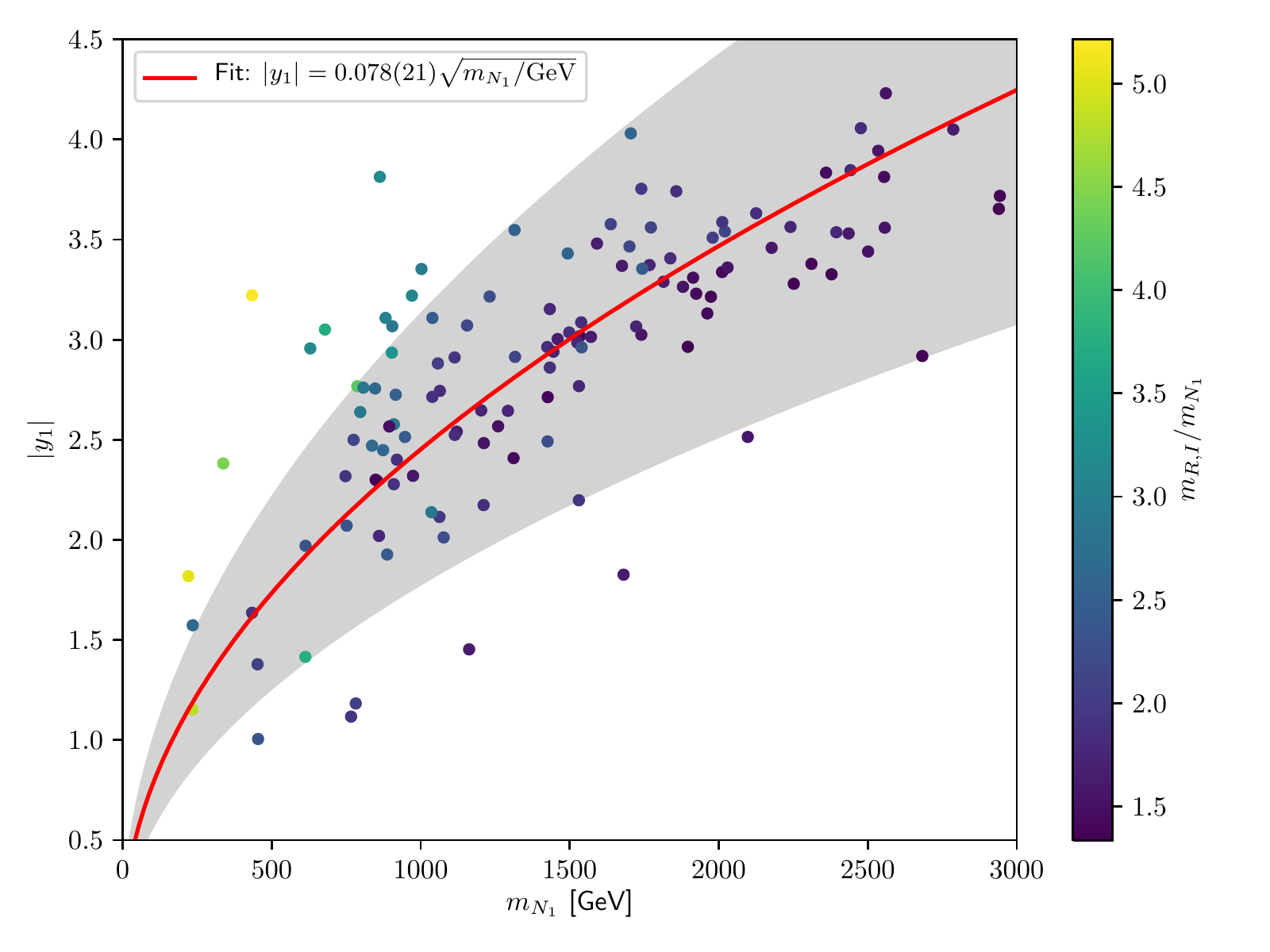}
\caption{Yukawa coupling of the lightest neutrino as a function of the DM mass.
  The ratio of the neutral scalar over the DM mass is given on the temperature scale.}
\label{fig:04}
\end{center}
\end{figure}

Furthermore, with $m_{\nu_1}/|\lambda_5|$ fixed, the Yukawas in Eq.\ (\ref{eq:8})
become correlated with the DM and scalar masses. Since the ratio of the
latter is in addition constrained by the relic density ($m_{R,I}/m_{N_1}\sim1.5$),
the leading term in Eq.\ (\ref{eq:8}) becomes proportional to $|y_1|^2/m_{N_1}$, which
allows us to fit this dependence in Fig.\ \ref{fig:04} at 90\% C.L.\ as
\beq
 |y_1|=\left\{\begin{array}{ll}
 (0.078\pm0.021) \ \sqrt{m_{N_1}/{\rm GeV}} & ({\rm NH})\\
 (0.081\pm0.012) \ \sqrt{m_{N_1}/{\rm GeV}} & ({\rm IH})\end{array}\right. ,
\eeq
the other fermions $N_{2,3}$ being significantly heavier. This implies, that if
the DM mass is known, we can predict its Yukawa coupling to the SM leptons.

\begin{figure}[t]
\begin{center}
\includegraphics[width=0.49\textwidth]{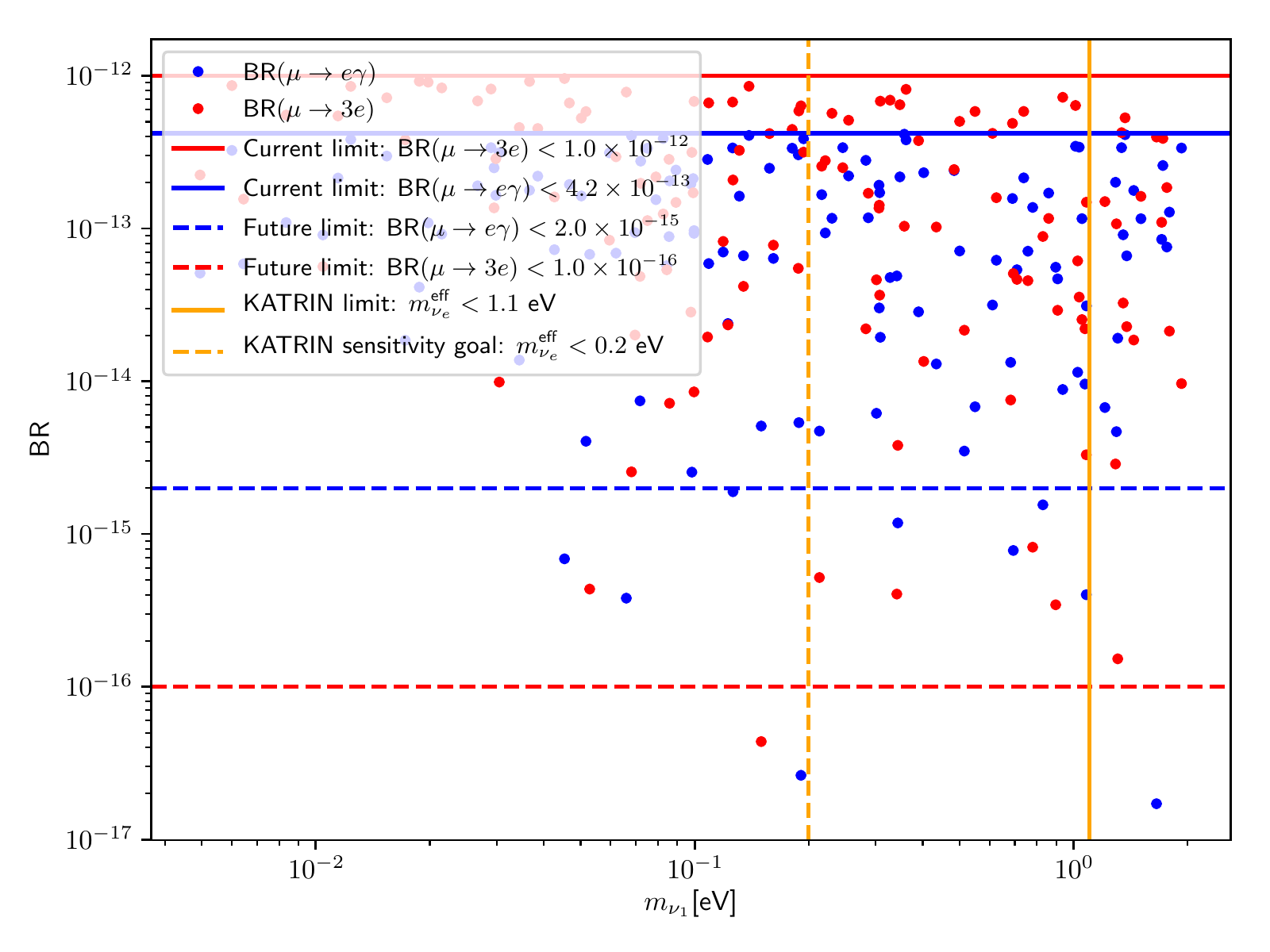}
\caption{Branching ratios of viable scotogenic models for the LFV processes $\mu\to e\gamma$
  (blue) and $\mu\to3e$ (red points), their current (full) \cite{TheMEG:2016wtm,Bellgardt:1987du}
  and future (dashed) \cite{Renga:2018fpd,Blondel:2013ia} experimental limits,
  and the current \cite{Aker:2019uuj} and future \cite{Drexlin:2013lha} KATRIN limits
  (yellow lines) on the (effective) electron neutrino mass.}
\label{fig:05}
\end{center}
\end{figure}

Our findings imply that the fermion DM parameter space of the scotogenic model can
be almost completely tested with LFV experiments and a measurement or limit on the direct
neutrino mass, as can be seen from Fig.\ \ref{fig:05}.
The current limit on BR$(\mu\to e\gamma)$ (blue) \cite{TheMEG:2016wtm} imposes
stronger bounds than the one for BR$(\mu\to 3e)$ (red) \cite{Bellgardt:1987du},
but this is expected to change soon \cite{Renga:2018fpd,Blondel:2013ia}.
Independently of the neutrino mass hierarchy, the models that survive
even these future tight constraints can be probed in an orthogonal way by new
limits or measurements of the absolute neutrino mass, if they reach indeed the
region of cosmologically favored values \cite{Aker:2019uuj,Drexlin:2013lha}.

In Tab.\ \ref{tab:1}, we show the input parameters for a typical benchmark point
\begin{table*}
\caption{Input parameters for a typical benchmark point at the KATRIN sensitivity limit (all masses in GeV).}
\label{tab:1}
\begin{tabular}{|ccc|ccccccccc|c|ccccc|}
\hline
$m_{N_1}$ & $m_{N_2}$ & $m_{N_3}$ & $y_{11}$ & $y_{12}$ & $y_{13}$ & $y_{21}$ & $y_{22}$ & $y_{23}$ & $y_{31}$ & $y_{32}$ & $y_{33}$ &
 $m_\eta$ & $\lambda_1$ & $\lambda_2$ & $\lambda_3$ & $\lambda_4$ & $\lambda_5$ \\
\hline
$1926$ & $3773$ & $3607$ & $-3.055$ & $-2.471$ & $0.183$ & $1.748$ & $-2.003$ & $2.038$ & $-1.180$ & $1.690$ & $2.673$ &
 $3371$ & $0.26$ & $0.5$ & $7.576$ & $6.060$ & $6.2\cdot10^{-10}$ \\
\hline
\end{tabular}
\end{table*}
at the KATRIN sensitivity limit (all masses in GeV). The corresponding
physical parameters, i.e.\ neutrino masses and mixings, relic density and
branching ratios/conversion rate for this NH benchmark point (all masses in
eV, angles in $^\circ$) are shown in Tab.\ \ref{tab:2}.
\begin{table*}
\caption{Neutrino masses and mixings, relic density and branching ratios/conversion rate for our NH benchmark point (all masses in eV, angles in $^\circ$).}
\label{tab:2}
\begin{tabular}{|c|cc|ccc|c|ccc|}
\hline
 $m_{\nu_1}$ & $\Delta m_{21}^2$ & $\Delta m_{31}^2$ & $\theta_{12}$ & $\theta_{23}$ & $\theta_{13}$ &
 $\Omega h^2$ & BR($\mu\to e\gamma$) & BR($\mu\to 3e$) & CR($\mu-e$,Ti) \\
\hline
 $0.2$ & $7.231\cdot10^{-5}$ & $2.444\cdot10^{-3}$ & $32.03$ & $44.28$ & $8.50$ &
 $0.120$ & $2.990\cdot10^{-13}$ & $8.321\cdot10^{-13}$ & $3.180\cdot10^{-13}$ \\
\hline
\end{tabular}
\end{table*}
We have checked that our results, in particular for the relic density, depend only
weakly on CP-violating phases, so that they have been neglected.
As one can see, this point fulfils all current constraints and is in the
sensitivity range of forthcoming upgrades of the LFV experiments.

\section{Summary and outlook}
\label{sec:5}

As we have demonstrated in this Letter, the parameter space of fermion DM in Ma's
scotogenic model is now severely constrained. In particular,
an electron neutrino mass measurement would allow to directly predict the
dark sector-Higgs boson coupling $\lambda_5$ and to test the complete parameter space
of the model in an orthogonal way to LFV, while a DM mass measurement would result
in a prediction for its Yukawa coupling to the SM leptons. This is due to the
strong mutual constraints inherent in the one-loop diagram (Fig.\ \ref{fig:01}) for
neutrino mass generation, which is topologically similar to a penguin diagram
mediating LFV and (when cut on the internal fermion line) to DM annihilation.
The correlations are absent for scalar DM, i.e.\ when the diagram is cut on the internal
scalar lines, since the scalar doublets can annihilate into weak gauge bosons. The case of
scalar-fermion coannihilation has been studied elsewhere \cite{Klasen:2013jpa}.

In particular, we extended previous findings \cite{Vicente:2014wga} in three different
directions, showing that the eigenvalues of the Yukawa couplings are actually of very
similar size with ratios close to unity; exposing the linear relation of $\lambda_5$
to the {\em absolute} electron neutrino mass as currently measured by KATRIN, while
previously only neutrino mass {\em differences} were taken into account; and
demonstrating the approximate square-root dependence of the Yukawa couplings on the
DM mass.

Our observations generalize to other scotogenic models such as those with triplet
fermions \cite{Ma:2008cu} and/or singlet-doublet scalars \cite{Farzan:2009ji},
where the neutrino mass matrices take forms similar to Eq.\ (\ref{eq:8}).
However, since the neutral components of electroweak triplets can annihilate
into gauge bosons, the Yukawas and LFV processes are generally smaller, so that collider
constraints must also be considered \cite{Cirelli:2014dsa,Dev:2015pga}.

 
\begin{acknowledgements}
  The authors thank C.\ Weinheimer for very helpful discussions and comments on the manuscript.
  This work has been supported by the DFG through the Research Training Network 2149
  ``Strong and weak interactions - from hadrons to dark matter''.
\end{acknowledgements}

\bibliography{bib}

\end{document}